\documentstyle[epsfig,here,12pt]{article}
\newcommand{\smallfrac}[2] {\mbox{$\frac{#1}{#2}$}}
\newcommand {\eqref} [1] {(\ref {#1})}
\newcommand {\beq} {\begin{equation}} 
\newcommand {\eeq} {\end{equation}}
 \newcommand {\ber}{\begin{eqnarray*}}
 \newcommand {\eer} {\end{eqnarray*}}
\newcommand {\bea}{\begin{eqnarray}}
 \newcommand {\eea} {\end{eqnarray}} 
\newcommand{\adss}{$AdS_5\times S^5\ $}
\newcommand{\Nfour} {${\cal N}=4\ $}
\newcommand{\Ntwo}{${\cal N}=2\ $}
\newcommand{\None}{${\cal N}=1\ $}
\newcommand{\drawsquare}[2]{\hbox{%
\rule{#2pt}{#1pt}\hskip-#2pt
\rule{#1pt}{#2pt}\hskip-#1pt
\rule[#1pt]{#1pt}{#2pt}}\rule[#1pt]{#2pt}{#2pt}\hskip-#2pt
\rule{#2pt}{#1pt}}
\newcommand{\Yfund}{\raisebox{-.5pt}{\drawsquare{6.5}{0.4}}}
\newcommand{\ztwo}{${\bf Z}_2\ $}

\def\Acknowledgements{\bigskip  \bigskip {\begin{center} \begin{large}
             \bf ACKNOWLEDGEMENTS \end{large}\end{center}}}

\begin{document}\begin{titlepage}
\rightline{TAUP-2577-99}
\rightline{\today}
\vskip 1cm
\centerline{{\Large \bf  Non-Supersymmetric Large $N$ Gauge Theories \newline
    }}
\centerline{{\Large \bf from Type 0 Brane Configurations }}
\vskip 1cm
\centerline{{\bf Adi Armoni} and {\bf Barak Kol}}

\begin{center}
\em School of Physics and Astronomy
\\Beverly and Raymond Sackler Faculty of Exact Sciences
\\Tel Aviv University, Ramat Aviv, 69978, Israel
\end{center}
\centerline{\bf armoni@post.tau.ac.il, barak@beauty.tau.ac.il}
\begin{abstract}
We use dyonic brane configurations of type 0 string theory to study large 
$N$ non-supersymmetric 4d gauge theories. The brane configurations
define theories  
similar to the supersymmetric ones which 
arise in type II. We find the non-SUSY analogues of \Ntwo and
\None. In particular we 
suggest
 new non-SUSY CFT's and a brane realization
of a non-SUSY Seiberg duality.
\end{abstract}
\end{titlepage}

\section{Introduction}
In recent years we learned that D-branes can be used as a
tool to study supersymmetric gauge theories\cite{HW} (for a recent
review and references see\cite{GK}). Using brane
configurations of type IIA/B string theory the Seiberg-Witten curves
of \Ntwo \cite{SW} were given a geometrical realization \cite{Witten1}, the
running of the coupling was related to the bending of the NS branes \cite{Witten1},
and \None Seiberg duality \cite{Seiberg} was realized by the exchange of the NS
supports \cite{EGK,EGKRS}. 

Though much understanding was gained for supersymmetric theories,
branes configurations failed to teach us about non-supersymmetric
gauge theories, since non-SUSY brane configurations are usually
not stable.

An important direction in the study of non-SUSY gauge theories is the
use of branes in the context of type 0 string theory\cite{KT1}. Type 0 string
theory is defined on the world sheet exactly
like type II,
except that a non-chiral GSO projection is performed. The resulting sectors
of the theory are the (NS-,NS-),(NS+,NS+) and a doubled set of R-R
fields. The low-energy fields of the theory are therefore a tachyon,
the bosonic (NS,NS) fields of type II and two-copies of R-R
fields. Accordingly two types of D-branes exists. We will refer to the
two kinds as 'electric' and 'magnetic', and to a pair of electric and
magnetic as 'dyonic'. 
The 'dyonic' combination of branes was called also 'untwisted' by
\cite{BG2}, as it belongs to the untwisted sector in
their description of type 0 as an orbifold of M theory. 

Klebanov and Tseytlin argued that in the background of D-branes, the
R-R flux will cure the tachyon instability and that the gauge theory on
the branes is perfectly ok. Using the electric D3 branes they
constructed an $SU(N)$ gauge theory with six adjoint scalars \cite{KT1} 
and studied its behavior in an AdS/CFT inspired way \cite{ads}. For
recent related works see \cite{KT3,type0}.

Another interesting construction uses a stack of $N$ coincident dyonic D3
branes\cite{KT3}. In this case each dyonic brane can be thought of as
a pair made of
an electric and a magnetic brane. The strings that connect the
electric-electric and the magnetic-magnetic branes yields a $SU^e(N)
\times SU^m(N)$ gauge theory with 6 magnetic adjoint scalars and 6 electric
adjoint scalars. In addition the strings that connect electric-magnetic branes
give rise to additional 4 bifundamental Weyl fermions in the $(N,\bar N)$
and another 4 bifundamental Weyl fermions in the $(\bar N,N)$
\cite{BG}. The
model has similarities to the orbifold models of Kachru and
Silverstein\cite{KS}. Although it cannot be obtained by a string
orbifold of type IIB, it can be viewed as a
field theory ``orbifold''
truncation (in the sense of ref.\cite{BJ}) of \Nfour $U(2N)$ Super
Yang-Mills\cite{NS}. An interesting remark is that although the theory
is non-SUSY it admits
a moduli space of vacua\cite{Zarembo,TZ}. 

The gravity solution of this configuration is an \adss
space with constant dilaton and zero tachyon. Therefore Klebanov and
Tseytlin interpreted this theory as a non-SUSY CFT\cite{KT3}.  
The fact that
the stringy solution of type 0 dyonic D3 branes is the same
as that of type II is not surprising. When we consider dyonic branes in
type 0
(i.e. identifying electric and magnetic branes), a solution with zero
tachyon and the same massless bosonic fields as in type II would
exist. The only additional conditions are large $N$ and small enough 't Hooft
coupling\cite{KT3}. The reason for the similarity is that in these cases, the
S-matrix of sphere amplitudes for fields common to type 0 and type II
coincide \cite{KT1}. Therefore the
low-energy effective action of type 0 is almost the same. The
differences are the existence of two R-R fields and the presence of the tachyon
which is coupled to other fields with even powers. As a result, we find
a solution of the equations of motion in which the tachyon is zero and
the two kinds of R-R fields are identified. This is exactly the type
II solution. An important remark is the necessity of large $N$. Higher
order contributions to the tree level action
(string loops) may contribute terms which do not have even powers of
tachyon, hence the zero tachyon solution cease to exist. However,
these contributions are suppressed in the large $N$ limit. Another
remark is that the R-R flux is expected to shift the tachyon mass to
positive values
only when the 't Hooft coupling is small enough. Therefore only in this
limit, the solution with zero tachyon is expected to be stable.

From the field theory point of view, the class of theories that we consider
here fall into the class of ``orbifold field theories''. It was shown
in ref.\cite{BJ}, and later generalized to the case of product groups
in \cite{Schmaltz},
that certain truncation of supersymmetric field theories would yield 
a non-supersymmetric gauge theory with exactly the 
same large N Green functions (in the untwisted sector). As we will
show, the type 0 truncation corresponds to such
an orbifold projection.

An important feature of these theories is that they exist only
for small enough values of the 't Hooft coupling (\cite{KT3} find
$\lambda < 100$ for the ``\Nfour'' theories).
 The orbifold non-SUSY theories inherit only a subset of the operators.
 In particular the operator $tr\ F_1^2 - tr\ F_2^2$ does not exist in
 the SUSY theory. This operator couples
to the tachyon in the bulk.
 As long as $\lambda$ is small enough, the two point
 function of this operator is non-tachyonic. However, for large values
 of $\lambda$ the dimension
of the operator becomes complex and the
 non-SUSY theory is not well defined.
 In the parent SUSY theory there is no such operator and therefore it is well defined for all
 values of $\lambda$. \footnote{We thank
  O. Aharony, A. Rajaraman and A. Tseytlin for clarifying this point.}

In this paper, we would like to generalize the D3 solution to the
more complicated cases when we include NS branes. Since the above
reasoning should hold in this case also, we expect that the
non-supersymmetric theory which lives on the analogous brane
configuration would share many of the properties of supersymmetric
one. In particular, in the large $N$ limit, these theories would
 have the same perturbative Green
functions and hence the same perturbative beta function. 
An important remark is that the
 validity of this approach is subjected to the assumption of 
non-perturbative
equivalence of the
non-SUSY untwisted sector and the parent.
The simplest example
is the \Ntwo brane configuration of type II.  Here we also expect to have
exactly flat dyonic directions and a mass spectrum 
in the dyonic sector
which is Bose-Fermi
degenerate, though non-SUSY. For the special case of
$N_f=2N_c$ we will have a large N non-SUSY CFT.
Another class of large N non-SUSY CFT's can be constructed from the
type 0 analogue of the 
brane-boxes configurations\cite{HZ,HSU}.
The type 0 analogue of the  \None configuration yields a
non-supersymmetric version of Seiberg duality. 
In particular, the type 0 theories are expected to maintain the phase
structure in the  $N_f/N_c$ axis.
These theories
are a special case of the theories 
considered in the past by Schmaltz who showed that the large N
``orbifold field theories'' of \None admits duality\cite{Schmaltz}. It
is interesting that in the present case the duality can be understood
also via branes.   

Some open questions are raised by this discussion. We discuss large
$N$ non-SUSY field theories which have the same dynamics as
supersymmetric ones. In particular these theories have the same number
of bosons and fermions. One wonders if there is any symmetry (not SUSY)
which is responsible for that. In the supersymmetric parent the
particle spectrum is divided to BPS and non-BPS. We expect the
distinction to carry over to the non-SUSY theory, such that the masses
of the ``BPS'' particles is given by a BPS-like formula. Is there an
intrinsic way to distinguish the two kinds?

The organization of the paper is as follows: In section 2 we describe
the type 0 string theories and their relation to the II string
theories. In section 3 we explain the orbifold truncation of field
theory and its relation to type 0 theories. Section 4 is devoted to
the study of the non-SUSY analogues of \Ntwo. In section 5 we
consider the non-SUSY version of Seiberg duality. We describe finite
non-SUSY models using brane boxes in section 6. Section 7 is devoted
to conclusions. 

\section{Type 0 String Theory}
Type II A/B string theories have non-supersymmetric analogues called
 type 0 A/B. The type 0 theories are constructed via a non-chiral GSO
 projection which keeps
 the following bosonic sectors
\vskip 0.5cm

type 0A:    $(NS-,NS-) \oplus (NS+,NS+) \oplus (R+,R-) \oplus (R-,R+)$

type 0B:    $(NS-,NS-) \oplus (NS+,NS+) \oplus (R+,R+) \oplus (R-,R-)$

\vskip 0.5cm

The tree level type 0B action from the $(NS+,NS+)$ sector
is exactly the same as the type IIB action\cite{KT1}
\begin{eqnarray}
S=-2\int d^Dx{\sqrt G}e^{-2\Phi}\left (R+4(\partial_n\Phi)^2
  -\frac{1}{12}H_{mnk}^2\right) \ ,
\end{eqnarray}
where $H_{mnk}$ is the field strength of the anti-symmetric two-form 
$B_{mn}$.
The action of the tachyon which comes from the $(NS-,NS-)$ sector is
\begin{eqnarray}
S=\int d^Dx{\sqrt G}e^{-2\Phi}\left(
  \smallfrac{1}{2}G^{mn}\partial_mT\partial_nT +\smallfrac{1}{2}m^2T^2
  \right) \ ,
\end{eqnarray}
where $m^2=\frac{2-D}{4\alpha'}$ is the mass of the tachyon.
The leading R-R terms in the action are
\begin{eqnarray}
S=\int d^Dx{\sqrt G} h_{(n+1)}(T)|F_{n+1}|^2 \ ,
\end{eqnarray}
where $F_{n+1}$ is the field strength of the {\em dyonic} R-R fields and $h_{(n+1)}(T)$
describes the coupling of the tachyon to the R-R fields.
Note that although the functions $h_{(n+1)}(T)$ are not known, 
symmetry arguments for the coupling to the 'electric' and 'magnetic'
forms require that these functions should be even in the dyonic case. 

It is therefore clear that a solution in a dyonic background with
$T=0$ is the same as the solution of the bosonic fields of type II
supergravity. While this statement is trivial in the tree level, it is
certainly wrong for $g_s>0$\cite{KT3}, since torus amplitudes generate
odd tachyon contributions to the action.
 Therefore similarities between
supersymmetric type II gauge theories and non-supersymmetric type 0
theories are expected to occur in the large $N$ limit, like the $D3$
type 0 case \cite{KT3}, the orbifold models of 
Kachru and Silverstein \cite{KS} and the non-supersymmetric version of
Seiberg \None duality which was considered by Schmaltz \cite{Schmaltz}.

Note also that in order to have a stable $T=0$ solution, we must shift
the tachyon mass. Since the function $h(T)$ contains a $T^2$ part, for
large enough R-R flux the tachyon mass-squared becomes positive. In the
background of a Dp brane, this condition translates into a requirement of
small 't Hooft coupling. 

\section{``Type 0'' projected field theories}
Given a brane configuration in type II 
string theory, one can
construct a new non-supersymmetric field theory by considering the
same brane configuration in type 0 with dyonic branes replacing the
type II D-branes.
This procedure can be given an intrinsic definition in field theory,
which we shall discuss now. The new field theory will be constructed by
a certain (orbifold) projection of a SUSY field theory. All
amplitudes with untwisted external legs of the non-SUSY theory will be identical, in the large $N$
limit, to amplitudes of the parent theory.

For ``\Nfour '' (the type 0 version of \Nfour) this was shown by
\cite{NS} who put the type 0 projection in a form suitable for the
more general methods of 
\cite{KS,BKV,BJ}. One starts with a
$U(2N_c)$ \Nfour theory and performs a \ztwo projection, keeping only the
invariant fields. The \ztwo is
embedded into the ${\bf Z}_4$ center of
the $SO(6)$ R-symmetry and its generator acts on the gauge group as
$\gamma = \left[ \begin{array}{c c } 1  & 0 \\ 0 & -1 \end{array}
\right]$ , where the entries are $N_c \times N_c$
blocks. The resulting spectrum is made of bosons in the adjoint of
$SU^e(N_c) \times SU^m(N_c)$ and fermions in the bifundamental. The
projection will preserve all amplitudes in the large N limit, since
if satisfies the condition that the matrix $\gamma$ be traceless. 

The generalization to theories with less SUSY and with the addition of
 flavor was discussed by
\cite{Schmaltz}. As SUSY QCD-like theories include matter with indices in a
$SU(2N_f)_L \times SU(2N_f)_R$ global symmetry, they are not explicitly
 included in the previous discussion that assumes
matter in the adjoint. One needs to embed the \ztwo
into the flavor group as well. 
The flavor indices are divided into electric and magnetic in the same
manner as the
color indices. All fields in QCD-like theories carry two indices which
 are either color-color or color-flavor. The projected matter content
 turns out to be made, again, 
 of bosons when both indices are of the same kind (say electric-electric), and
fermions for magnetic-electric, where now each index can be either
 color or flavor. In \cite{Schmaltz} it was checked and
proven that the large N amplitudes are preserved. 
An important remark is that the non-SUSY theory contains operators, such
as $tr\ F_1^2-tr\ F_2^2$, which do not exist in the parent theory. These
operators may become tachyonic (their anomalous dimension may become
complex) and therefore the non-SUSY theory will be sick.

\section{The ``\Ntwo'' Brane Configuration}

Consider the type 0 analogue of an \Ntwo theory with gauge group
$SU(N_c)$ and $N_f$ matter hypermultiplets. We refer to it as a
``\Ntwo'' theory. To determine the matter
content one can either use the ``type 0'' projection of a $SU(2N_c)$
theory with $2N_f$ hypers, or read it off the appropriate brane
configuration, as we shall. In type II, the relevant brane
configuration is a stack of $N_c$ "color" D4 branes extending between
two NS5
 supports, and $N_f$ semi-infinite "flavor" D4's
ending on the supports. Strings stretched between the color branes
give the vectors in the adjoint, while the strings between the color
and flavor branes give hypers in the fundamental. In passing to type
0, we replace both color and flavor D4's by dyonic D4's - a mix of an equal
number of electric
and magnetic branes. 

\begin{figure}[H]
  \begin{center}
\mbox{\kern-0.5cm
\epsfig{file=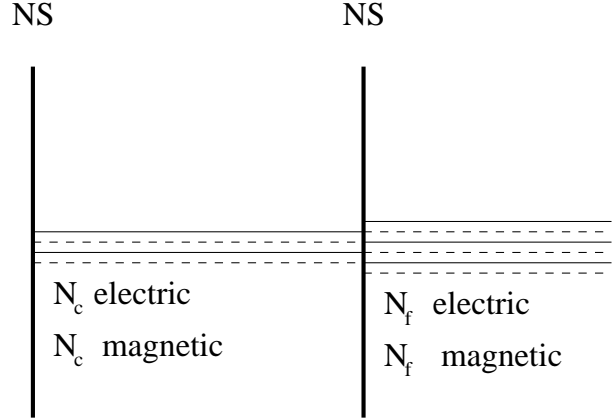,width=8.0true cm,angle=0}}
\label{N2_fig}
  \end{center}
\caption{The ``\Ntwo `` type 0 brane configuration. Each dyonic brane
  can be viewed as a pair of electric and magnetic branes.}
\end{figure}

One can now read the matter content after
recalling that electric-electric (EE) or magnetic-magnetic (MM)
strings are bosons, while electric-magnetic (EM) strings are
fermions. The matter content is summarized in table (\ref{table_N=2}) below.

\begin{table}[H]
\begin{displaymath}
\begin{array}{l c@{ } c@{ } c@{ } c@{ } c}
 & \multicolumn{1}{c@{\times}}{SU^e(N_c)}
& \multicolumn{1}{c@{\times}}{SU^m(N_c)}
& \multicolumn{1}{c@{\times}}{SU^e(N_f)}
& \multicolumn{1}{c@{}}{SU^m(N_f)} \\
\hline
\rm {vector} & adj. & 1 & 1 & 1 \\
\rm{vector} & 1 & adj. & 1 & 1 \\
2\ \rm {scalars}  & adj. & 1 & 1 & 1 \\
2\ \rm {scalars}  &  1 & adj. & 1 & 1 \\
2\ \rm{fermions} &  \Yfund & \overline{\Yfund} & 1 & 1 \\
2\ \rm {fermions} &   \overline{\Yfund} & \Yfund & 1 & 1 \\ 
\hline
\rm {complex\ scalar} &  \Yfund & 1 & \overline{\Yfund} & 1 \\
\rm {complex\ scalar} &   \overline{\Yfund} & 1 & \Yfund & 1 \\
\rm {complex\ scalar} &  1 & \Yfund & 1 & \overline{\Yfund}  \\
\rm {complex\ scalar} &   1 & \overline{\Yfund} & 1&  \Yfund \\
\rm {Weyl\ fermion} &  \Yfund & 1 &  1 & \overline{\Yfund}  \\
\rm {Weyl\ fermion} &   \overline{\Yfund} & 1 & 1 & \Yfund \\
\rm {Weyl\ fermion} &  1 & \Yfund  & \overline{\Yfund} & 1 \\
\rm {Weyl\ fermion} &   1 & \overline{\Yfund} &  \Yfund & 1 

\end{array}
\end{displaymath}
\caption{The matter content of the ``\Ntwo'' theory}
\label{table_N=2}
\end{table}

It is immediately seen that the ``\Ntwo'' theory has the same 1-loop
beta function as its parent \Ntwo theory:
\beq
\beta = 2N_c-N_f.
\eeq
In comparison with the original \Ntwo theory we have the same number
of fields in the fundamental, while we traded the gauginos in the
adjoint with fermions in the bifundamental (and its complex
conjugate). But these two representations have the same group theoretic
factor $N_c=T({\rm adj})=2N_c T({\rm fund}), T(R) \delta ^{ab}=tr_R
[T^aT^b]$. However, other group theoretical factors of the two representations
do differ, and so will higher loop computations (for finite $N_c$).

One can hope for simplifications in the large $N_c$ limit. In addition
to keeping the 't Hooft coupling $\lambda= g_{YM}^2 N_c$, fixed, we
fix the flavor ratio $\nu_f=N_f/N_c$ and the scale of mass of the W's
(we are assuming that the large $N_c$ theory has a moduli space).   

For ``\Nfour'' theories is was proven \cite{NS,BKV,LNV,BJ} that in the
large $N$ limit, it
has the same untwisted
amplitudes as the parent \Nfour - one views both
$U(N_c)$ ``\Nfour'' and $U(N_c)$, \Nfour as projections of $U(2N_c)$, \Nfour,
and hence correlation functions of fields which are shared by both
theories (namely, untwisted bosons in the adjoint) must be the same. As discussed in the
previous section, we find that this is the case for the ``\Ntwo''
theories as well.
 Supporting evidence
comes from the gravitational background of the type 0
brane configuration - as the type 0 background is identical to the one
in type II, and given the field theory - gravity correspondence\cite{ads}, one
expects the theories to have the same amplitudes.
 This implies some exact results for these non-SUSY large $N_c$
 theories: the $N_f=2N_c$ theory
would be exactly conformal. 
For any $\nu_f$ the theories would have exact 
dyonic flat
directions, leading to an (infinite dimensional) moduli space. For
``\Nfour'' a moduli space is expected to exist as a consequence of the no force between
dyonic 3 branes in type 0 \cite{KT1,TZ}. And Finally, the mass
of dyonic
``BPS'' particles would be expected to be free of corrections.  
In particular, the masses of the W's are expected to be the same as
of the supersymmetric theory, since the relevant Green functions are
the same.   

\section{``\None'' and Seiberg Duality}

Let us consider now the type 0 analogue of \None SQCD. The
supersymmetric (electric) theory consists of a $SU(N_c)$ vector multiplet and two
chiral $SU(N_f)$ multiplets. The brane realization of the theory is
similar to the \Ntwo configuration and it is obtained by rotating one
of the NS5 branes such that half of the supersymmetries are
broken. Passing to the type 0 theory , the
matter content is summarized in the following table

\begin{table}[H]
\begin{displaymath}
\begin{array}{l c@{ } c@{ } c@{ } c@{ }c@{ } c@{ } c}
 & \multicolumn{1}{c@{\times}}{SU^e(N_c)}
& \multicolumn{1}{c@{\times}}{SU^m(N_c)}
& \multicolumn{1}{c@{\times}}{SU^e(N_f)}
& \multicolumn{1}{c@{\times}}{SU^m(N_f)}
& \multicolumn{1}{c@{\times}}{SU^e(N_f)}
& \multicolumn{1}{c@{}}{SU^m(N_f)} \\
\hline
\rm {vector} & adj. & 1 & 1 & 1 & 1 & 1\\
\rm {vector} & 1 & adj. & 1 & 1 & 1 & 1 \\
\rm {fermion} &  \Yfund & \overline{\Yfund} & 1 & 1 & 1 & 1\\
\rm {fermion} &   \overline{\Yfund} & \Yfund & 1 & 1 & 1 & 1 \\ 
\hline
 \rm {scalar} &  \Yfund & 1 & \overline{\Yfund} & 1 & 1 & 1 \\
 \rm {scalar} &   \overline{\Yfund} & 1 & 1 & 1 & \Yfund & 1 \\
 \rm {scalar} &  1 & \Yfund & 1 & \overline{\Yfund}  & 1 & 1\\
 \rm {scalar} &   1 & \overline{\Yfund} & 1 & 1 & 1 &  \Yfund \\
 \rm {fermion} &  \Yfund & 1 &  1 & \overline{\Yfund} & 1 & 1 \\
 \rm {fermion} &   \overline{\Yfund} & 1 & 1 & 1 & 1 & \Yfund \\
 \rm {fermion} &  1 & \Yfund  & \overline{\Yfund} & 1& 1 & 1\\
 \rm {fermion} &   1 & \overline{\Yfund} & 1 & 1 &  \Yfund & 1 \\

\end{array}
\end{displaymath}
\caption{The content of the electric ``\None'' theory}
\label{table:N=1}
\end{table}

One may consider also a magnetic theory which is based on
$SU(N_f-N_c)$ gauge group, $N_f$ flavors and elementary Meson
field. The type 0 analogue of this theory is 
\begin{table}[H]
\begin{displaymath}
\begin{array}{l c@{ } c@{ } c@{ } c@{ }c@{ } c@{ } c}
 & \multicolumn{1}{c@{\times}}{SU^e({\tilde N_c})}
& \multicolumn{1}{c@{\times}}{SU^m({\tilde N_c})}
& \multicolumn{1}{c@{\times}}{SU^e(N_f)}
& \multicolumn{1}{c@{\times}}{SU^m(N_f)}
& \multicolumn{1}{c@{\times}}{SU^e(N_f)}
& \multicolumn{1}{c@{}}{SU^m(N_f)} \\
\hline
\rm {vector} & adj. & 1 & 1 & 1 & 1 & 1\\
\rm {vector} & 1 & adj. & 1 & 1 & 1 & 1 \\
 \rm {fermion} &  \Yfund & \overline{\Yfund} & 1 & 1 & 1 & 1\\
 \rm {fermion} &   \overline{\Yfund} & \Yfund & 1 & 1 & 1 & 1 \\ 
\hline
 \rm {scalar} &  \overline{\Yfund} & 1 & \Yfund & 1 & 1 & 1 \\
 \rm {scalar} &   \Yfund & 1 & 1 & 1 & \overline{\Yfund} & 1 \\
 \rm {scalar} &  1 & \overline{\Yfund} & 1 & \Yfund  & 1 & 1\\
 \rm {scalar} &   1 & \Yfund & 1 & 1 & 1 &  \overline{\Yfund} \\
 \rm {fermion} &  \overline{\Yfund} & 1 &  1 & \Yfund & 1 & 1 \\
 \rm {fermion} &   \Yfund & 1 & 1 & 1 & 1 & \overline{\Yfund} \\
 \rm {fermion} &  1 & \overline{\Yfund}  & \Yfund & 1& 1 & 1\\
 \rm {fermion} &   1 & \Yfund & 1 & 1 &  \overline{\Yfund} & 1 \\
\hline
\rm {scalar} & 1 & 1 & \overline{\Yfund} & 1 & \Yfund & 1 \\
\rm {scalar} & 1 & 1 & 1 & \overline{\Yfund} & 1 & \Yfund \\
\rm {fermion} & 1 & 1 & \overline{\Yfund} & 1 & 1 & \Yfund \\
\rm {fermion} & 1 & 1 & 1 & \overline{\Yfund} & \Yfund & 1
\end{array}
\end{displaymath}
\caption{The content of the magnetic ``\None'' theory. ${\tilde N_c} = N_f-N_c$.}
\label{table:N'=1}
\end{table}

Seiberg showed that the electric and the magnetic theories describes
the same IR physics. The duality was realized geometrically via an exchange
of the NS branes\cite{EGK,EGKRS}. We claim that the large N 
(keeping $\nu_f=N_f/N_c$ fixed)
 type 0 electric and magnetic theories are
also dual. The field theory reasonings, for a wider class of theories,
was given in \cite{Schmaltz}.

The branes gives a simple evidence that \None duality holds. One may
simply repeat the same steps of \cite{EGK}. The dyonic branes dynamics of
type 0 is identical to that of type II and therefore starting with the
electric theory one would end with the magnetic theory.

The field theory arguments in favor of the duality are based
on\cite{Schmaltz}. Since the Green functions of the untwisted sector, in the large N, are
identical, it follows that the type 0 field theory has the same phase
 structure and
dynamics as the type II field theory and therefore duality between the
supersymmetric theories implies duality between the non-SUSY ones.

In particular it it easy to check that the one loop beta function of
the electric theory type 0 theory is
\beq
\beta = 3N_c-N_f,
\eeq
exactly as of the \None theory. Moreover, the field theory analysis of
\cite{BJ,Schmaltz} suggests that a conformal window exists for
${3\over 2} N_c < N_f < 3N_c$. Another simple check is the anomaly.
Both the type 0 electric theory and magnetic theory have the same
global anomalies. The reason is \cite{Schmaltz} that the anomalies are
calculated by planar diagrams which are the same in the non-SUSY
theory and
its parent. For instance,
\bea
 & & SU^3(N_f) \sim N_c \\
 & & SU^2(N_f) U_B (1) \sim N_c \nonumber
\eea

Thus, though the theory is not supersymmetric we have a good
understanding of the IR physics and the phase structure. 

\section{Brane Boxes and CFT's}

 Brane boxes suggest a nice realization of finite four dimensional
supersymmetric gauge theories via type IIB string theory\cite{HZ,HSU}.
 Let us review, briefly, the construction of these
configurations. The reader is referred to \cite{HSU} for a comprehensive
discussion. Consider a set of $N$ D5 branes along 012346 directions, 
a set of parallel NS branes along 012345 directions and additional set of
parallel NS' branes along 012367 directions. The models that we
consider here are on a torus in the 46 directions. The resulting theory is,
generically, \None gauge theory in four dimensions.  The theories are
described in the following figure

\begin{figure}[H]
  \begin{center}
\mbox{\kern-0.5cm
\epsfig{file=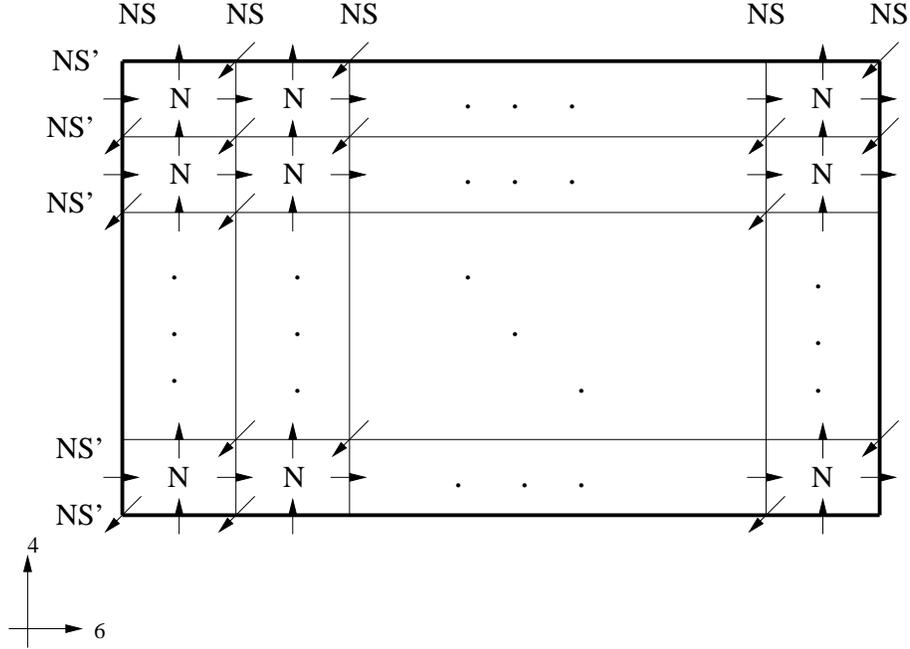,width=12.0true cm,angle=0}}
\label{box_fig}
  \end{center}
\caption{Finite brane-boxes theories. There are $k\times k'$
  boxes. The drawn 46 plane is a torus.}
\end{figure}

Let us denote the number of horizontal boxes by $k$ and the number of
vertical boxes by $k'$. Each box describes an $SU(N)$ gauge group
which interacts with the other boxes via open strings. The allowed
directions of strings (which are drawn as arrows) are north, south,
east, west and two of the diagonals: northwest and
southeast. Note that the northeast and the southwest
diagonal arrows are not allowed\cite{HZ}. Each triangle of arrows form
a contribution to the superpotential $W$. The sign of the contribution to the
superpotential is dictated by the handedness of the triangle.
Thus a generic finite model is an $SU(N)\times ... \times
SU(N)$ ($kk'$ times) gauge theory with bifundamental matter. 
In addition there might be additional matter due to the intersection of
the NS-NS' branes. This issue was addressed lately in the framework of
'diamonds' models \cite{AKLM}, where it was pointed out that the
intersection singularity should be blown up as give rise to a more
involved diamond interactions. However, in the specific
case that we present here, namely a zero size diamond, 
we will adopt the conservative view that the field content
should be the same as of \cite{HSU}.  
The NS/NS' branes
do not bend since there are the same number of D branes in each side of
them. In addition there is a superpotential. The form of the
superpotential together with the fact that the NS/NS' branes do not
bend guarantee that these models are indeed finite.

The case of $k=k'=1$ is special. In this case the supersymmetry is enhanced
to \Nfour. When $k=1,k'>1$ the supersymmetry is enhanced to \Ntwo. The
generic case $k,k'>1$ corresponds to \None supersymmetry.

These models can be constructed in type 0 also. In the large $N$ we
will have non-supersymmetric CFT's which are analogues to the above
\Nfour, \Ntwo and \None models. The rule of constructing the type 0
gauge theory is described in section 3. Put $N$ electric and $N$
magnetic D-branes in each box. Accordingly the gauge group will be
$SU^e(N)\times SU^m(N)\times ... \times SU^e(N) \times SU^m(N)$ ($kk'$
electric and $kk'$ magnetic groups). Strings that connect
electric-electric or magnetic-magnetic branes are bosons whereas
strings that connect electric-magnetic branes are fermions. 

Let us consider the special example of $k=k'=2$. The supersymmetric
model consist of a gauge group $SU(N) \times SU(N) \times SU(N) \times
SU(N)$ with vector like matter in the bifundamental of each of the two
groups. 

The analogous type 0 theory contains eight gauge groups. The vectors (``gluons'')
are in the adjoints of the gauge groups, the ``gauginos'' are in the
bifundamental of $SU^e(N)\times SU^m(N)$ pairs which originate from a
same gauge group, i.e. eight Weyl
fermions. The matter consist of six complex scalars (``squarks'') in $SU^e(N)\times SU^e(N)$
and six complex scalars in $SU^m(N)\times SU^m(N)$. In addition there are
twelve Weyl fermions (``quarks'') which belong to $SU^e(N) \times
SU^m(N)$. The matter fields originate from distinct $SU(N)$ groups.

Again, it is easy to see that the one loop beta function is zero. The
proofs of \cite{BJ,Schmaltz} guarantee that the theory is indeed
finite. 

These non-SUSY CFT's are new in the sense that they cannot be
constructed by an orbifold of type IIB\cite{KS,LNV}.    

\section{Conclusions}
In this paper we constructed various non-supersymmetric large N gauge
theories which were suggested to share many properties of their
supersymmetric parents.
In particular, we found CFT's, a special version of Seiberg duality
(which was considered first in\cite{Schmaltz}), a degenerate mass
spectrum and a moduli space of vacua.

The construction of non-supersymmetric brane configurations suggests
many directions of research. We didn't consider theories in higher or
lower dimensions than four and we didn't consider branes in the presence of
orientifolds (orientifolds 
of type 0 theories were discussed in
\cite{sagnotti}). It seems that many results which were obtained via type
II brane configurations can be easily copied to the type 0 case.

Finally, we would like to refer the reader to phenomenological aspects
of orbifold gauge theories\cite{pheno}. It was argued that the standard model may be
included in ``\Nfour'' like theories. This scenario suggests that the
underlying theory is not supersymmetric but conformal.
\Acknowledgements

We thank Ofer Aharony, Vadim Kaplunovsky, Arvind Rajaraman, Jacob
Sonnenschein and  Shimon Yankielowicz for useful discussions.

\end{document}